\documentclass[seceq,preprint]{ptptex}

\usepackage{graphicx}

\makeatletter
\def\simleq{\mathrel{\mathpalette\gl@align<}}
\def\simgeq{\mathrel{\mathpalette\gl@align>}}
\def\gl@align#1#2{\lower.6ex\vbox{\baselineskip\z@skip\lineskip\z@
     \ialign{$\m@th#1\hfill##\hfil$\crcr#2\crcr\sim\crcr}}}
\makeatother

\newcommand{\nn}{\nonumber\\}

\newcommand{\Pu}{p_{\uparrow}}

\newcommand{\Nu}{n_{\uparrow}}
\newcommand{\Nd}{n_{\downarrow}}

\title{
Exploring Three-Nucleon Forces in Lattice QCD%
}

\author{
Takumi~\textsc{Doi,}$^{1,}$%
\footnote{
E-mail: doi@ribf.riken.jp
}%
\footnote{
Present address:
Theoretical Research Division, Nishina Center, RIKEN, Wako 351-0198, Japan
}
Sinya~\textsc{Aoki,}$^{1,2}$
Tetsuo~\textsc{Hatsuda,}$^{3,4,5}$
Yoichi~\textsc{Ikeda,}$^{6}$
Takashi~\textsc{Inoue,}$^{7}$
Noriyoshi~\textsc{Ishii,}$^{2}$
Keiko~\textsc{Murano,}$^{5}$
Hidekatsu~\textsc{Nemura}$^{8,}$%
\footnote{
Present address:
Center for Computational Sciences, University of Tsukuba, Tsukuba 305-8577, Japan
}
and
Kenji~\textsc{Sasaki}$^{2}$ \\
(HAL QCD Collaboration)\\
\bigskip
\includegraphics[width=0.35\textwidth]{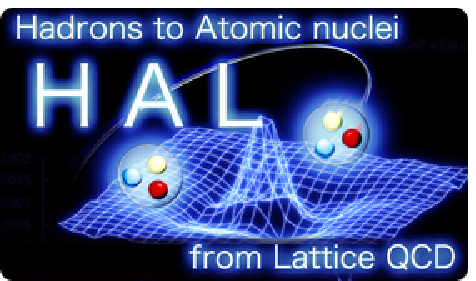}
}

\inst{
$^1$%
{Graduate School of Pure and Applied Sciences, University of Tsukuba,
 Tsukuba 305-8571, Japan}
\\
$^2$%
{Center for Computational Sciences, University of Tsukuba, Tsukuba 305-8577, Japan}
\\
$^3$%
{Department of Physics, The University of Tokyo, Tokyo 113-0033, Japan}
\\
$^4$%
{IPMU, The University of Tokyo, Kashiwa 277-8583, Japan}
\\
$^5$%
{Theoretical Research Division, Nishina Center, RIKEN, Wako 351-0198, Japan}
\\
$^6$%
{Department of Physics, Tokyo Institute of Technology,
Meguro, Tokyo 152-8551, Japan}
\\
$^7$%
{Nihon University, College of Bioresource Sciences, Fujisawa 252-0880, Japan}
\\
$^8$%
{Department of Physics, Tohoku University, Sendai 980-8578, Japan}
}

\abst{
Three-nucleon forces (3NF) are investigated
from two-flavor lattice QCD simulations.
We utilize the Nambu-Bethe-Salpeter (NBS) wave function
to determine two-nucleon forces (2NF) and 
3NF in the same framework.
As a first exploratory study,
we extract 3NF in which three nucleons are aligned linearly with an 
equal spacing. This is the simplest  geometrical configuration 
which reduces the huge computational cost of calculating the  NBS wave function.
Quantum numbers of the three-nucleon system are chosen to be
$(I, J^P)=(1/2,1/2^+)$ (the triton channel).
Lattice QCD simulations are performed using 
$N_f=2$ dynamical clover fermion configurations
at the lattice spacing of $a = 0.156$ fm on a $16^3\times 32$ lattice
with a large quark mass corresponding to $m_\pi= 1.13$ GeV.
We find repulsive 3NF at short distance in the triton channel.
Several sources of systematic errors are also discussed.
}

\PTPindex{200, 234, 232, 164}  

\begin{document}

\maketitle

\section{Introduction}
\label{sec:intro}

The nuclear force is an essential input 
 in studying the  nuclear many-body problems. In particular, 
 the ``realistic nuclear potentials''
between two nucleons (2N), which can successfully reproduce
the vast amount of the 2N scattering data,
 have been used in  ab-initio nuclear calculations.
It turns out, however, that these two-nucleon forces (2NF)
generally underestimate the experimental binding energies of 
light nuclei~\cite{Kamada:2001tv, Pieper:2007ax};
 this indicates the necessity of
  three-nucleon forces (3NF). 
 High precision deuteron-proton  elastic scattering experiments at
  intermediate energies have also shown a clear indication of
   3NF~\cite{Sekiguchi:2011ku, Sekiguchi:2008zzc}.

There also exist various  phenomena
in nuclear physics and astrophysics where 3NF
may play an important role. Some examples include
(i) the cross section of the
  backward scattering angles in 
 nucleus-nucleus elastic scattering~\cite{Furumoto:2009zz},
(ii) the anomaly in the oxygen isotopes near the neutron drip-line~\cite{Otsuka:2009cs}, and 
(iii) the nuclear equation of state (EoS)
  at high density  relevant to the physics of neutron stars~\cite{Akmal:1998cf}.
  Universal short-range repulsion for three baryons (nucleons and hyperons) 
  has also been suggested in relation to the maximum mass of 
   neutron stars with hyperon core~\cite{Nishizaki:2002ih, Takatsuka:2008zz}.

Despite of its phenomenological importance,
microscopic understanding  of 3NF is  still limited.
Pioneered by Fujita and Miyazawa~\cite{Fujita:1957zz},
the long range part of 3NF has been modeled
 by the two-pion exchange (2$\pi$E)~\cite{Coon:2001pv},
particularly with the $\Delta$-resonance excitation.
This 2$\pi$E-3NF component is known to have 
an attractive nature at long distance.
An additional repulsive component of 3NF at short distance is 
often introduced in a purely phenomenological way~\cite{Pieper:2001ap}.
 An approach based on  the chiral effective field theory (EFT)
 is quite useful to   classify and parametrize 
 the  two-, three- and more-nucleon forces~\cite{Weinberg:1992yk, Epelbaum:2008ga, Machleidt:2011zz}.
%
%
While unknown low-energy constants cannot be determined 
within its framework but are obtained only 
by the fitting to the experimental data,
the obtained parameters are used in 
nuclear physics calculations~\cite{Epelbaum:2011md}.

To go beyond phenomenology,
it is most desirable to determine 3NF
directly
from the fundamental degrees of freedom (DoF), the quarks and the gluons,
on the basis of quantum chromodynamics (QCD).
In this letter,
we make a first exploratory study of 
 first-principle lattice QCD calculation of 3NF.%
\footnote{
We note here that a completely different approach 
based on holographic QCD
leads to a nuclear matrix model whose solution gives 
repulsive 3NF at short distance~\cite{Hashimoto:2010ue}.
}

As for the 2NF from lattice QCD,
 an approach based on  the 
 Nambu-Bethe-Salpeter (NBS) wave function 
 has been  proposed~\cite{Ishii:2006ec, Aoki:2009ji}. 
 Resultant (parity-even) 2NF in this approach
are found to have attractive wells at long and medium
distances  and central repulsive cores at short distance.
The method has been  extended also to the
hyperon-nucleon (YN) and hyperon-hyperon (YY) interactions%
~\cite{Nemura:2008sp, Inoue:2010hs, Sasaki:2010bi,Inoue:2010es}.
From the numerical point of view, 
the extension of the above  methodology to three-nucleon (3N) systems
is rather challenging due to
 huge number of DoF of the NBS wave function for 3N systems.
Also, we need to identify genuine 
3NF by subtracting out the 2NF in both parity-even channel and parity-odd
channel.  The latter channel has not yet been extracted 
 in the present lattice QCD simulations.

In this paper,
as a first exploratory study,
 we consider the triton channel, $I=1/2$, $J^P= 1/2^+$,
as the quantum numbers of the 3N system.
In this case, we can determine 3NF 
even without the knowledge of parity-odd 2NF.
In order to reduce the computational cost, 
we study 3NF 
 through a configuration of three nucleons aligned linearly with an 
 equal spacing, which is the simplest geometrical configuration (a linear setup) 
for the 3N system.

 Lattice QCD simulations are carried out
 with $N_f=2$ dynamical clover fermion configurations.
Preliminary accounts of this study have been 
given in Refs.~\citen{Doi:2010yh, Doi:baryons2010}.
It is in order here to mention that
lattice QCD simulations for 
three- and four- baryon systems have been also  performed 
in Refs.~\citen{Yamazaki:2009ua, Beane:2009gs}.
They, however, focus on the 
energies of the multi-baryon systems,
 and extracting 3NF, which could be used in 
  future ab-initio nuclear physics calculations, is currently beyond their scope.

This paper is organized as follows.
In Section~\ref{sec:formulation:3NF},
we describe the basic formulation to study 3N systems,
as well as its challenges.
In Section~\ref{sec:formulation:linear},
we develop a framework to obtain 3NF in the triton channel
using only parity-even 2NF.
The advantage of the ``linear setup'' for 
the three-dimensional (3D) coordinate configuration is also discussed.
In Section~\ref{sec:setup:linear},
we explain our lattice QCD setup,
and numerical results of 3NF
are presented in Section~\ref{sec:results:linear}.
Section~\ref{sec:summary} is devoted to summary 
and concluding remarks.
In Appendix~\ref{sec:eff_2N},
we present a complementary study using 
effective 2NF in 3N systems.

\section{Formulation for three-nucleon systems}
\label{sec:formulation:3NF}

To begin with, 
we briefly review the framework for 
the calculation of 2N potentials~\cite{Aoki:2009ji}.
We consider the following effective Schr\"odinger equation,
\begin{eqnarray}
-\frac{\nabla^2}{2\mu} \psi_{2N}(\vec{r})
+ \int d\vec{r}' U_{2N}(\vec{r},\vec{r}') \psi_{2N}(\vec{r}')
= E_{2N} \psi_{2N}(\vec{r}) ,
\label{eq:Sch_2N}
\end{eqnarray}
where $\mu = m_N/2$ and $E_{2N}$ 
are the reduced mass and ground-state energy of the 2N system
in the center-of-mass frame, respectively,
and $U_{2N}(\vec{r},\vec{r}')$ is the non-local energy-independent 2N potential.
The equal-time NBS wave function $\psi_{2N}(\vec{r})$ is 
extracted from the four-point correlator as
\begin{eqnarray}
\label{eq:4pt_2N}
G_{2N} (\vec{r},t-t_0)
&\equiv& 
\frac{1}{L^3}
\sum_{\vec{R}}
\langle 0 |
          (N(\vec{R}+\vec{r}) N (\vec{R}))(t)\
\overline{(N' N')}(t_0)
| 0 \rangle , \\
&\xrightarrow[t \gg t_0]{}& A_{2N} \psi_{2N} e^{-E_{2N}(t-t_0)} , 
\quad
A_{2N} = \langle E_{2N} | \overline{(N' N')} | 0 \rangle , \\
\label{eq:NBS_2N}
\psi_{2N}(\vec{r}) &\equiv& \langle 0 | N(\vec{r}) N(\vec{0}) | E_{2N}\rangle ,
\end{eqnarray}
where 
$|E_{2N}\rangle$ denotes the ground state vector of the 2N system,
$N$ ($N'$) the nucleon operator in the sink (source),
and 
spinor/flavor 
indices are suppressed for simplicity.
In the practical calculation,
we perform the derivative expansion for the non-locality of the potential~\cite{okubo-marshak},
\begin{eqnarray}
U_{2N}(\vec{r},\vec{r}') =
\left[ V_C(r) + V_T(r) S_{12} + V_{LS}(r) \vec{L}\cdot \vec{S} + {\cal O}(\nabla^2) \right]
\delta(\vec{r}-\vec{r}') ,
\label{eq:Okubo_Marshak}
\end{eqnarray}
where $V_C$, $V_T$ and $V_{LS}$ are the central, tensor and spin-orbit potentials, respectively.
In Ref.~\citen{Murano:2011nz}, the validity of this expansion is examined,
and it is shown that the leading terms, $V_C$ and $V_T$, dominate the potential at low energies.

The extension to 3N systems is described as follows.
We consider the NBS wave function $\psi_{3N}(\vec{r},\vec{\rho})$ 
extracted from the six-point correlator as
\begin{eqnarray}
\label{eq:6pt_3N}
G_{3N} (\vec{r},\vec{\rho},t-t_0) 
&\equiv& 
\frac{1}{L^3}
\sum_{\vec{R}}
\langle 0 |
          (N(\vec{x}_1) N(\vec{x}_2) N (\vec{x}_3))(t) \
\overline{(N'       N'        N')}(t_0)
| 0 \rangle ,\ \ \ \\
&\xrightarrow[t \gg t_0]{}& A_{3N} \psi_{3N} (\vec{r},\vec{\rho}) e^{-E_{3N}(t-t_0)} ,
A_{3N} = \langle E_{3N} | \overline{(N' N' N')} | 0 \rangle , \ \ \ \\
\label{eq:NBS_3N}
\psi_{3N}(\vec{r},\vec{\rho}) &\equiv& 
\langle 0 | N(\vec{x}_1) N(\vec{x}_2) N(\vec{x}_3) | E_{3N}\rangle ,
\end{eqnarray}
where
$E_{3N}$ and $|E_{3N}\rangle$ denote
the energy and the state vector of the 3N ground state, respectively, and
$\vec{R} \equiv ( \vec{x}_1 + \vec{x}_2 + \vec{x}_3 )/3$,
$\vec{r} \equiv \vec{x}_1 - \vec{x}_2$, 
$\vec{\rho} \equiv \vec{x}_3 - (\vec{x}_1 + \vec{x}_2)/2$
the Jacobi coordinates.
Assuming that the NBS wave function 
has a proper asymptotic behavior at outer non-interacting regions,
as was proven in the case of 2N systems%
~\cite{Aoki:2009ji, Luscher:1990ux, Lin:2001ek, Aoki:2005uf, Ishizuka:2009bx},
we consider the following 
Schr\"odinger equation of the 3N system 
with the derivative expansion of the potentials,
\begin{eqnarray}
%
\biggl[ 
- \frac{1}{2\mu_r} \nabla^2_{r} - \frac{1}{2\mu_\rho} \nabla^2_{\rho} 
+ \sum_{i<j} V_{2N} (\vec{r}_{ij})
+ V_{3NF} (\vec{r}, \vec{\rho})
\biggr] \psi_{3N}(\vec{r}, \vec{\rho})
= E_{3N} \psi_{3N}(\vec{r}, \vec{\rho}) , 
\label{eq:Sch_3N}
\end{eqnarray}
where
$V_{2N}(\vec{r}_{ij})$ with $\vec{r}_{ij} \equiv \vec{x}_i - \vec{x}_j$
denotes the 2NF between $(i,j)$-pair,
$V_{3NF}(\vec{r},\vec{\rho})$ the 3NF,
$\mu_r = m_N/2$, $\mu_\rho = 2m_N/3$ the reduced masses.

3NF can be determined as follows.
We first calculate 
$\psi_{3N}(\vec{r}, \vec{\rho})$ for all $\vec{r}$, $\vec{\rho}$
and obtain the total potential of the 3N system 
through Eq.~(\ref{eq:Sch_3N}).
We also perform (separate) lattice simulations
for genuine 2N systems and obtain 
all necessary $V_{2N}(\vec{r}_{ij})$.
We then extract $V_{3NF}(\vec{r},\vec{\rho})$
by subtracting $\sum V_{2N}(\vec{r}_{ij})$ from the total potential. 
The extension to four- and more-nucleon forces can be immediately understood.
Note that potentials determined in this way reproduce the 
energy of the system by construction.

Important remark is that 3NF are always determined in combination with
2NF, and 3NF alone do not make too much sense.
The comparison between lattice 3NF and phenomenological 3NF can be done 
only at a qualitative level.
Rather, our purpose is to determine two-, three-, (more-) nucleon forces 
systematically, and provide them as a consistent set. 
%

In the practical calculation with the above procedure, 
the computational cost is enormous:
Note that adding one more nucleon to the system corresponds to
adding three (valence) quarks, each of which has
three color and four Dirac spinor DoF.
In addition,
the number of diagrams 
to be calculated in the Wick contraction
tends to diverge with a factor of $N_u ! \times N_d !$,
where $N_u$ ($N_d$) are numbers of up (down) quarks in the system.
We develop several techniques to reduce these computational costs.
For instance, we take advantage of symmetries (such as isospin symmetry)
to reduce the number of the Wick contractions.
We also 
employ the non-relativistic limit for the nucleon operator in the source
to reduce the number of spinor DoF.
Similar techniques are
(independently) developed in Ref.~\citen{Yamazaki:2009ua}.
Nevertheless, 
the simulation cost remains quite enormous, compared
to the current computational resources,
so we explore an efficient way to restrict the spacial DoF of $\vec{r}$ and $\vec{\rho}$,
as will be explained in the next section.

Even after we manage to calculate $\psi_{3N}(\vec{r},\vec{\rho})$,
the identification of genuine 3NF remains nontrivial.
In general, 
in order to subtract the contributions from $V_{2N}$ in Eq.~(\ref{eq:Sch_3N}),
we need $V_{2N}$ in both of parity-even and parity-odd channels,
since a 2N-pair inside the 3N system 
could be either of positive or negative parity.
In practice, however, 
the parity-odd $V_{2N}$
have not been obtained yet
in lattice QCD.
Note that the partial wave expansion cannot be performed here,
since we calculate only restricted DoF of $\vec{r}$ and $\vec{\rho}$
to reduce the computational cost.
In the next section, 
we establish a framework to extract 3NF 
without referring to parity-odd 2NF.


\section{Identification of three-nucleon forces and the linear setup for three-nucleon locations}
\label{sec:formulation:linear}

We study the 3N system in the triton channel, 
and consider Eq.~(\ref{eq:Sch_3N}) with fixed 
3D coordinate configurations of 3N.
In particular, we choose 
small $|\vec{r}|$, $|\vec{\rho}|$ configurations.
This corresponds to locating 3N close to each other,
and the 3NF effect is expected to be enhanced.

To proceed,
the lack of the information of parity-odd 2NF on the lattice
prevents us from identifying 3NF, as was explained in Section~\ref{sec:formulation:3NF}.
In order to resolve this issue,
we consider the following channel,
\begin{eqnarray}
\psi_S \equiv
\frac{1}{\sqrt{6}}
\Big[
-   \Pu \Nu \Nd + \Pu \Nd \Nu               
                - \Nu \Nd \Pu + \Nd \Nu \Pu 
+   \Nu \Pu \Nd               - \Nd \Pu \Nu
\Big]  ,
\label{eq:psi_S}
\end{eqnarray}
which is anti-symmetric
in spin/isospin spaces 
for any 2N-pair.
Combined with the Pauli-principle,
it is automatically guaranteed that
any 2N-pair couples with even parity only.
Therefore, 
parity-odd 2NF vanish
in $\langle \psi_S | H | \psi_{3N} \rangle$,
where $H$ is the Hamiltonian of the 3N system,
and we can extract 3NF unambiguously 
using only parity-even 2NF.
Note that no assumption on the choice of 3D-configuration of $\vec{r}$, $\vec{\rho}$
is imposed in this argument,
and we can take advantage of this feature
for 3NF calculations with various 3D-configuration setup,
which will be our subject in future simulations.

As noted in Section~\ref{sec:formulation:3NF},
it is important to
restrict the spacial DoF of $\vec{r}$ and $\vec{\rho}$ efficiently,
in order to reduce the computational cost of $\psi_{3N}(\vec{r},\vec{\rho})$.
In this paper,
we propose to use 
the linear setup with $\vec{\rho}=\vec{0}$. 
In this setup,
three nucleons are aligned linearly with equal spacings of 
$r_2=|\vec{r}_2|$ with $\vec{r}_2 \equiv \vec{r}/2$.

The advantage of the linear setup is understood as follows.
Because of $\vec{\rho}=\vec{0}$, the third nucleon is attached
to $(1,2)$-nucleon pair with only S-wave.
Considering that the total 3N quantum numbers are $I=1/2, J^P=1/2^+$ (triton channel), 
the wave function can be completely spanned by
only three bases, 
labeled
by the quantum numbers of $(1,2)$-pair as
$^{2S+1}L_J=$ $^1S_0$, $^3S_1$, $^3D_1$.
Therefore, the Schr\"odinger equation 
can be 
simplified to
the $3\times 3$ coupled channel equations
with the bases of 
$\psi_{\,^1\!S_0}$, $\psi_{\,^3\!S_1}$, $\psi_{\,^3\!D_1}$.
The reduction of the dimension of bases 
is expected to improve the S/N as well.
It is worth mentioning that
considering the linear setup is not an approximation:
Among various geometric components of 
the wave function of the ground state, 
we calculate the (exact) linear setup component
as
a convenient choice to study 3NF.
While we can access only a part of 3NF from it,
we plan to extend the calculation to more general geometries
step by step,
toward the complete determination of the full 3NF.

We examine
the explicit form of the potential matrix of 2NF.
At the leading order 
of the derivative expansion,
2NF is 
written in terms of
central potentials $V_C^{IS}$ and tensor potentials $V_T^{IS}$
with isospin $I$ and spin $S$,
\{%
$V_C^{00}$, 
$V_C^{10}$, 
$V_C^{01}$, 
$V_C^{11}$, 
$V_T^{01}$,
$V_T^{11}$%
\},
where the label ``$2N$'' is omitted for simplicity.
An explicit calculation gives us 
\newpage
%
%
%
%
%
%
\begin{eqnarray}
\label{eq:pot_3H_sym}
\lefteqn{
\sum_{i<j} V_{2N} (\vec{r}_{ij})\Big|_{\vec{\rho}=\vec{0}}
=}\nn
&&
\left(
\renewcommand{\arraystretch}{1.8}
\begin{array}{c|c|c}
                                   V_C^{10}(r_2) +             V_C^{01}(r_2)                    & 
                       \frac{1}{2} V_C^{10}(r_2) - \frac{1}{2} V_C^{01}(r_2)                    & 
- 2 V_T^{01}(r_2) + 2 V_T^{01}(r) \\                                                              
+ \frac{1}{2} V_C^{10}(r) + \frac{1}{2} V_C^{01}(r) & 
- \frac{1}{2} V_C^{10}(r) + \frac{1}{2} V_C^{01}(r) & \qquad\qquad   \\[2mm] \hline 
                           \frac{1}{2} V_C^{10}(r_2) - \frac{1}{2} V_C^{01}(r_2)                     & 
\frac{3}{4} V_C^{00}(r_2) + \frac{1}{4} V_C^{10}(r_2) + \frac{1}{4} V_C^{01}(r_2)                     & 
 V_T^{01}(r_2) - 3 V_T^{11}(r_2) + 2 V_T^{01}(r) \\                                                     
- \frac{1}{2} V_C^{10}(r) + \frac{1}{2} V_C^{01}(r) & 
+ \frac{3}{4} V_C^{11}(r_2) + \frac{1}{2} V_C^{10}(r) + \frac{1}{2} V_C^{01}(r) &   \\[2mm] \hline 
- 2 V_T^{01}(r_2) + 2 V_T^{01}(r)  &                                                           
  V_T^{01}(r_2) - 3 V_T^{11}(r_2) + 2 V_T^{01}(r) &                                             
  \frac{1}{2} V_C^{01}(r_2) + \frac{3}{2} V_C^{11}(r_2) - V_T^{01}(r_2)                  \\     
 &  & - 3 V_T^{11}(r_2) + V_C^{01}(r) - 2 V_T^{01}(r)  
\end{array}
\right) , \nn
\end{eqnarray}
%
%
%
%
%
%
where we span the spaces with the rotated bases given by
$(\psi_S, \psi_M, \psi_{\,^3\!D_1})^T$,
where
$\psi_S$ in Eq.~(\ref{eq:psi_S}) is shown to be
$\psi_S = \frac{1}{\sqrt{2}} ( - \psi_{\,^1\!S_0} + \psi_{\,^3\!S_1} )$,
and
$\psi_M \equiv \frac{1}{\sqrt{2}} ( + \psi_{\,^1\!S_0} + \psi_{\,^3\!S_1} )$.
Note that 
neither of 
parity-odd 2N potentials,
$V_C^{00}$, $V_C^{11}$, $V_T^{11}$, appear in the 
first row/column 
in Eq.~(\ref{eq:pot_3H_sym}), 
which is a consequence of the discussion above.

\section{Setup of lattice QCD simulations}
\label{sec:setup:linear}

We employ
$N_f=2$ dynamical 
configurations
with mean field improved clover fermion 
and 
renormalization-group improved
gauge action
generated by CP-PACS Collaboration~\cite{Ali_Khan:2001tx}.
We use
598 configurations at
$\beta \equiv 6/g^2 = 1.95$,
where the lattice spacing of 
$a^{-1} = 1.269(14)$ GeV
is determined from the rho meson mass~\cite{Ali_Khan:2001tx},
and 
the lattice size of $V = L^3 \times T = 16^3\times 32$
corresponds to
(2.5 fm)$^3$ box in physical spacial size.
For $u$, $d$ quark masses, 
we take the hopping parameter at the unitary point
as
$\kappa_u = \kappa_d = \kappa_{ud} = 0.13750$.
The masses of pion, nucleon and Delta are obtained as 
$m_\pi a = 0.8934(4)$,
$m_N a = 1.694(1)$ and
$m_\Delta a = 1.820(2)$
respectively,
corresponding to
$m_\pi = 1.13$ GeV, 
$m_N = 2.15$ GeV,
$m_\Delta = 2.31$ GeV.

We use the wall quark source with Coulomb gauge fixing,
and periodic (Dirichlet) boundary condition is imposed
in spacial (temporal) direction.
Due to the enormous computational cost,
we can perform the simulations only at a few sink time slices.
Looking for the range of sink time where the ground state saturation is achieved,
we carry out preparatory simulations for effective 2NF in the 3N system
in the triton channel.
We find that effective 2NF at different sink time slices
agree with each other as long as $(t-t_0)/a \geq 7$.
Being on the safer side,
we decide to perform linear setup calculations at $(t-t_0)/a =$ 8 and 9.
For details of the study of effective 2NF, see Appendix~\ref{sec:eff_2N}.

In order to enhance the statistics,
we perform the measurement at 32 source time slices for each configuration,
and the forward and backward propagations are averaged.
The results from both of total angular momentum $J_z=\pm 1/2$ 
are averaged as well.
The error is estimated by the jackknife method with
the bin size of 13, and the auto-correlation is found to be 
small by comparing the results with those from different bin sizes.
We carry out the simulation
at eleven physical points of the distance $r_2 = r/2$ with the linear setup.
More specifically,
we take 
$\vec{r}_2/a = $
$(0,0,0)$,
$(1,0,0)$,
$(1,1,0)$,
$(1,1,1)$,
$(2,0,0)$,
$(1,1,2)$,
$(2,2,0)$,
$(3,0,0)$,
$(2,2,2)$,
$(3,3,0)$,
$(3,3,3)$,
and additional lattice points obtained by cubic rotation from the above.
These points are chosen so that 
they can make a good sampling of $r_2$ as a whole,
and the number of lattice points in cubic rotation is small.
Furthermore, 
in order to suppress the finite volume artifact,
points with relatively large $r_2$ ($r_2 \simgeq 0.5$ fm)
are located in off-axis direction.

In the calculation of the four- and six-point correlators
(Eqs.~(\ref{eq:4pt_2N}) and (\ref{eq:6pt_3N})),
we have a freedom to choose any nucleon interpolating operator in
the sink and source, in so far as it (strongly) couples to a nucleon state.
Since a NBS wave function is defined through a sink operator $N$,
a different potential is obtained from a different operator $N$.
Note, however, that 
physical observables calculated from these different potentials, 
such as phase shifts and binding energies,
are unique by construction~\cite{Aoki:2009ji}.
In this sense, 
one can consider that 
choosing $N$ corresponds to choosing the ``scheme'' 
to define the potential~\cite{Aoki:2009ji, Aoki:2010ry}.
We choose $N$ so that 
the non-locality of the obtained potential is small,
achieving a reasonable S/N for the potential at the same time.
To this end,
we have found that the standard nucleon (sink) operator 
$N=N_{std} \equiv \epsilon_{abc}(q_a^T C \gamma_5 q_b) q_c$ 
satisfies these criteria in the 2N study~\cite{Murano:2011nz}.
We therefore employ the same sink operator in the 3N study as well,
so that 2NF and 3NF are determined 
on the same footing.
We still have an additional freedom to choose a source nucleon operator $N'$.
Since this choice does not affect the NBS wave functions nor the potentials,
our main concern here is the reduction of the computational cost.
In the 3N study,
we employ the non-relativistic limit operator,
$N'=N_{nr} \equiv \epsilon_{abc}(q_a^T C \gamma_5 P_{nr} q_b) P_{nr} q_c$
with $P_{nr} = (1+\gamma_4)/2$,
so that the spinor DoF is reduced and 
the faster computation is achieved.
In the 2N case,
we perform explicit lattice simulations
for both of $N'=N_{nr}$ choice
and $N'=N_{std}$ choice.
Theoretically, these potentials should agree at $t \rightarrow \infty$,
and we observe that these are actually consistent 
within statistical fluctuations at our time region of $(t-t_0)/a = 8$ and $9$.
Hereafter, we employ the source operator of $N_{nr}$ in the 2N study as well.
When extracting 3NF, we expect that this choice leads to a better
cancellation between 2NF and 3NF
for the statistical fluctuations as well as 
systematic error by the excited state contaminations.

\section{Numerical results of lattice QCD simulations}
\label{sec:results:linear}

\begin{figure}[t]
\begin{center}
\includegraphics[width=0.78\textwidth]{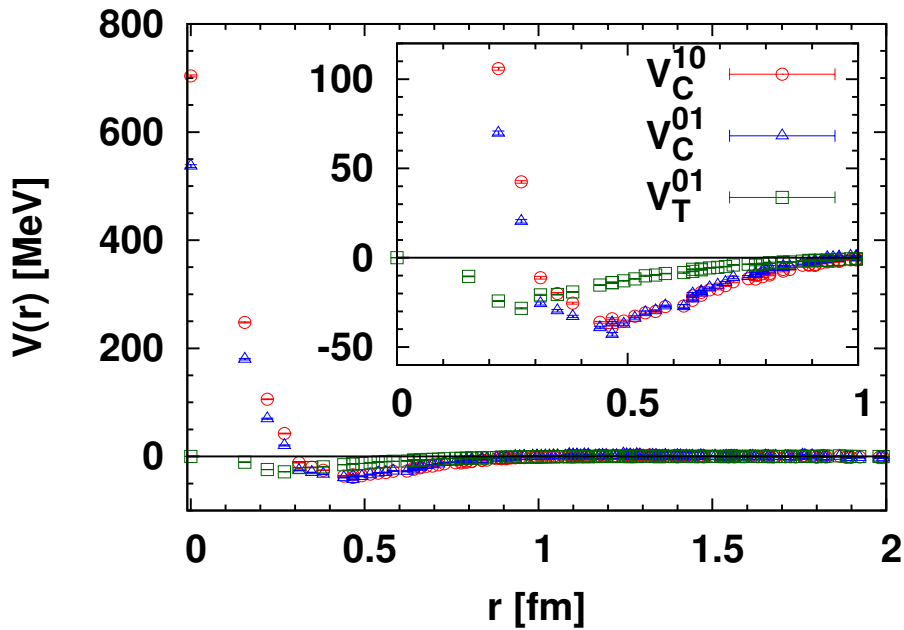}
\caption{
\label{fig:pot_2N}
Parity-even 2NF obtained 
at $(t-t_0)/a = 8$.
$V_C^{10}$ denotes the central force
in $^1S_0$ channel, and
$V_C^{01}$ and $V_T^{01}$
the central and tensor forces
in $^3S_1$-$^3D_1$ channel, respectively.
}
\end{center}
\end{figure}

We first show numerical results for 2N systems.
Shown in Fig.~\ref{fig:pot_2N} are
the parity-even 2NF of
$V_C^{10}$, $V_C^{01}$ and $V_T^{01}$ obtained at $(t-t_0)/a = 8$,
where we have checked that 
the sink time dependence is small as long as $(t-t_0)/a \geq 7$. 
$V_C^{10}$ is obtained from the single channel analysis of the $^1S_0$ state,
while
$V_C^{01}$ and $V_T^{01}$ are obtained from 
the $2\times 2$ coupled channel analysis in the $^3S_1$-$^3D_1$ state.
Each potential consists of the kinetic term (Laplacian term) and 
the energy term in Eq.~(\ref{eq:Sch_2N}).
The latter is estimated to be 
$E_{2N}^{10} \simeq -3$ MeV and $E_{2N}^{01} \simeq -3$ MeV 
in $^1S_0$ and $^3S_1$-$^3D_1$ channels, respectively,
using the condition that the potential approaches zero at large $r$
(the method called ``from $V$'' in Ref.~\citen{Aoki:2005uf}).

We observe central repulsive cores at short distance,
$r \simleq$ 0.2-0.3 fm,
and attractive wells at middle through long distances,
$0.3 \simleq r \simleq 0.8$ fm.
The qualitative features are same as our previous lattice simulations%
~\cite{Ishii:2006ec, Aoki:2009ji},
while quantitative strength and range of potentials 
are associated with the particular lattice setup in this calculation,
e.g., quark masses and a lattice spacing.

Let us turn to the results for the 3N system in the triton channel.
Each of $\psi_S$, $\psi_M$, $\psi_{\,^3\!D_1}$
is extracted by the projection in spin and flavor spaces.
Simultaneously, 
the projection to the $A_1^+$ representation of the cubic group
(which corresponds to the S-wave projection in the continuum limit)
is performed on $\psi_S$ and $\psi_M$,
while the subtraction of the $A_1^+$ representation
is performed on $\psi_{\,^3\!D_1}$.
In $\psi_{\,^3\!D_1}$, there are various $Y_{l=2,m}$ spherical harmonics components.
We find that they are generally consistent with each other
by making a division with a proper $Y_{2,m}$.
This indicates that the contamination from higher waves in $\psi_{\,^3\!D_1}$ are negligible,
and 
that 
the artifact due to the breaking of the rotational symmetry to the cubic symmetry
is small.%
\footnote{
Small discrepancy is observed only at $r_2/a = \sqrt{2}$,
which signals a splitting between $E$ and $T_2$ representations.
This artifact, however, makes negligible effect on results of 3NF.
}

\begin{figure}[t]
\begin{center}
\includegraphics[width=0.78\textwidth]{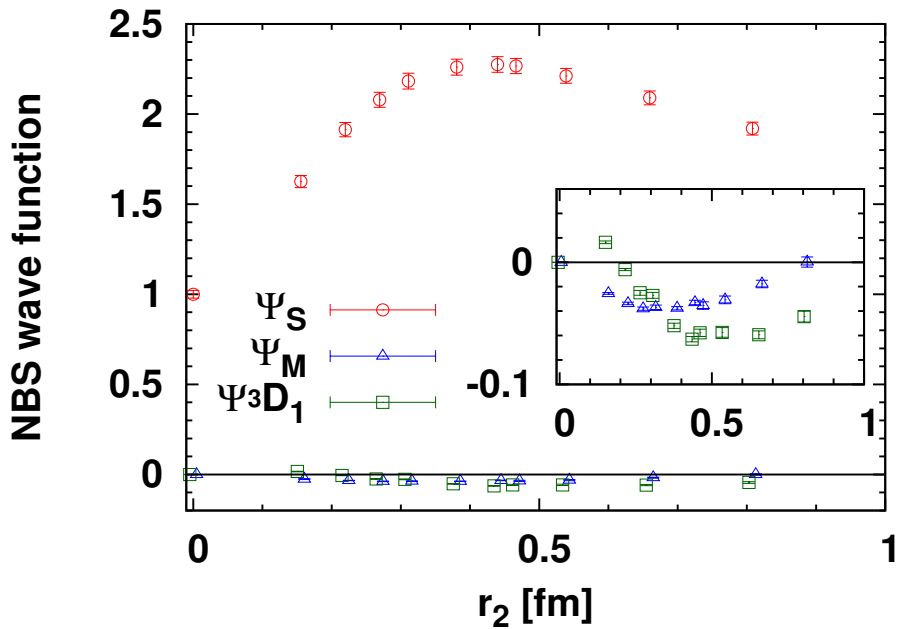}
\caption{
\label{fig:wf}
3N wave functions 
at $(t-t_0)/a=8$.
Circle (red), triangle (blue), square (green) points denote
$\psi_S$, $\psi_M$, $\psi_{\,^3\!D_1}$, respectively.
$r_2=r/2$ (with offset for visibility)
is
the distance between the center and edge
in the linear setup.
}
\end{center}
\end{figure}

In Fig.~\ref{fig:wf},
we plot
the radial part of each wave function of
$\psi_S$, $\psi_M$, $\psi_{\,^3\!D_1}$ 
obtained at $(t-t_0)/a = 8$.
Here, we normalize the wave functions
by the center value of $\psi_S(r_2=0)$.
What is noteworthy in Fig.~\ref{fig:wf} is that
the wave functions are obtained with good precision.
This is quite nontrivial,
since the S/N generally gets worse
when the system of concern is composed of more quarks~\cite{Lepage:1989hd}.
Furthermore, by fixing the 3D-configuration of 3N,
each data point is obtained from effectively
fewer number of statistical samplings.
This is in contrast to, e.g., calculations of the energy,
where there exists an effective gain on number of statistics 
by summations over the lattice volume for each nucleon.

In Fig.~\ref{fig:wf},
we observe that 
$\psi_S$ overwhelms the wave function,
and $\psi_M$, $\psi_{\,^3\!D_1}$ are much smaller
by one to two orders of magnitude.
This indicates that 
higher partial wave components in $\psi_S$
are also strongly suppressed,
and 
the wave function
is completely dominated by 
the component with which all three nucleons are in S-wave.
We consider that 
this is mainly originated by the large quark mass employed in this calculation.
For instance, $\psi_{\,^3\!D_1}$ component is expected to be suppressed 
due to smaller tensor forces with heavier pions.
In fact, in the case of spin-triplet 2N system,
lattice simulations with several quark masses
explicitly show that the D-wave component is 
smaller at larger quark masses~\cite{Aoki:2009ji}.

\begin{figure}[t]
\begin{center}
\includegraphics[width=0.78\textwidth]{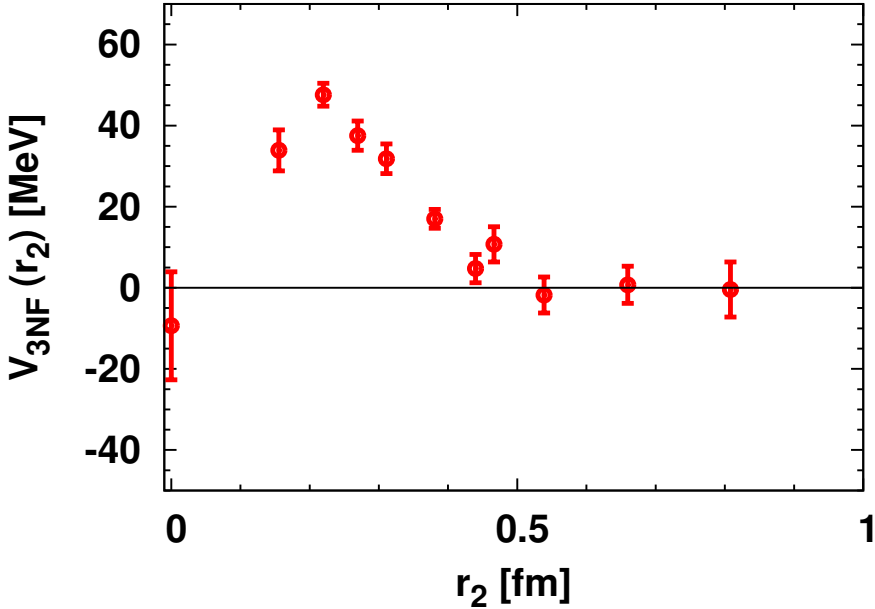}
\caption{
\label{fig:TNR}
The effective scalar-isoscalar 3NF 
in the triton channel with the linear setup
obtained at $(t-t_0)/a = 8$.
}
\end{center}
\end{figure}

We determine 3NF
by subtracting $V_{2N}$
from the total potentials in the 3N system.
As is explicitly shown in Eq.~(\ref{eq:pot_3H_sym}),
we have only one channel, 
$\langle \psi_S | H | \psi_{3N} \rangle$,
which is free from the parity-odd $V_{2N}$.
Correspondingly, we can determine 
only the sum of the 3NF matrix elements,
$
\langle \psi_S | V_{3NF} | \psi_{3N} \rangle
= 
\langle \psi_S | V_{3NF} | \psi_S \rangle +
\langle \psi_S | V_{3NF} | \psi_M \rangle +
\langle \psi_S | V_{3NF} | \psi_{^3D_1} \rangle
$,
or one type of spin/isospin functional form for 3NF. 
In this paper,
3NF are
effectively represented in 
a scalar-isoscalar functional form.
This form is an efficient representation,
since $\psi_S$ overwhelms the wave function and thus
$|\langle \psi_S | V_{3NF} | \psi_S \rangle|
\gg |\langle \psi_S | V_{3NF} | \psi_M \rangle|, |\langle \psi_S | V_{3NF} | \psi_{^3D_1} \rangle|$
is expected.
Note also a scalar-isoscalar functional form is 
often employed for the 
short-range part of 3NF in phenomenological models~\cite{Pieper:2001ap}.

\begin{figure}[t]
\begin{center}
\includegraphics[width=0.78\textwidth]{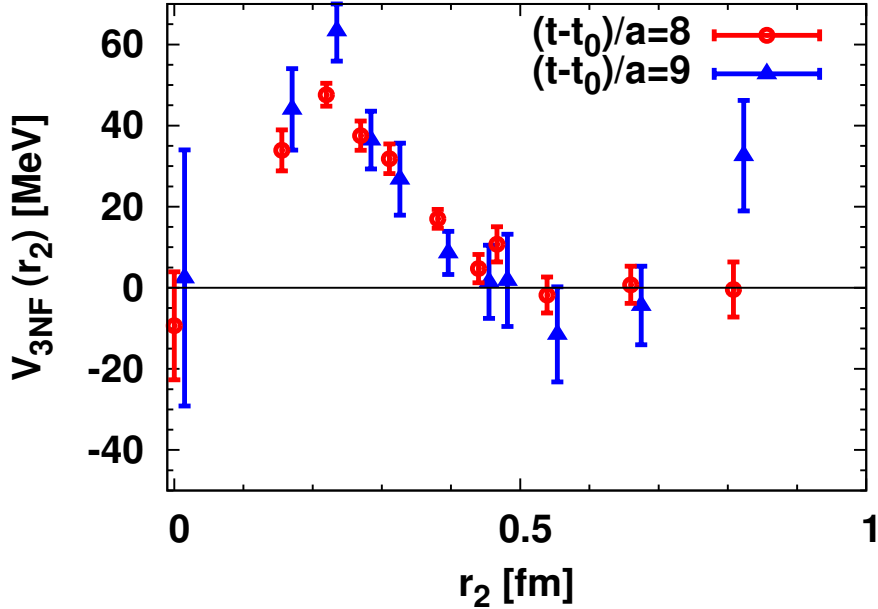}
\caption{
\label{fig:TNF_t-dep}
3NF as explained in Fig.~\ref{fig:TNR}.
Circle (red) points are obtained at $(t-t_0)/a = 8$,
and triangle (blue) points 
are obtained
at $(t-t_0)/a = 9$.
}
\end{center}
\end{figure}

When examining 3NF,
we have to include $r_2$-independent shift by energies,
in addition to kinematic terms (Laplacian part in Eqs.~(\ref{eq:Sch_2N}) and (\ref{eq:Sch_3N})).
As was done for the 2NF,
we employ the ``from $V$'' method~\cite{Aoki:2005uf},
and obtain $E_{3N} \simeq -3$ MeV from the 
long-range behavior of the effective 2NF in the 3N system
(see Appendix~\ref{sec:eff_2N} for details.)
Combined with the energies of 2N,
the total energy shift for the 3NF 
amounts to $\delta_E = E_{3N} - 3/2 E_{2N}^{10} - 3/2 E_{2N}^{01}$ $\simeq 5$~MeV.
Note that the precise determination of energies is still a challenging issue,
and we estimate that $\delta_E$ suffers from $\simleq 10$ MeV systematic error.
This uncertainty, however, does not affect the following discussions much,
since $\delta_E$ merely serves as an overall offset.
In Fig.~\ref{fig:TNR}, we plot the results
for the effective scalar-isoscalar 3NF at $(t-t_0)/a = 8$.
In order to check the dependence on the sink time slice,
we compare 3NF from $(t-t_0)/a =$ 8 and 9 
in Fig.~\ref{fig:TNF_t-dep}.
While the results with $(t-t_0)/a=9$ suffer from quite large errors, 
they agree with each other within statistical fluctuations.

Fig.~\ref{fig:TNR} shows that
3NF are small at the long distance region of $r_2$.
This is in accordance with the suppression
of 2$\pi$E-3NF by the heavy pion.
At the short distance region, on the other hand,
we observe an indication of repulsive 3NF.
Note that a repulsive short-range 3NF component
is phenomenologically required 
to explain the properties of high density matter.
Since multi-meson exchanges are strongly suppressed 
by the large quark mass on a lattice,
the origin of this short-range 3NF may be attributed to the 
quark and gluon dynamics directly.
In fact, we recall that the short-range repulsive (or attractive) cores
in the generalized baryon-baryon potentials 
are systematically calculated in lattice QCD in the flavor SU(3) limit~\cite{Inoue:2010hs},
and the lattice results are found to be well explained 
from the viewpoint of the Pauli exclusion principle in the quark level~\cite{Inoue:2010hs, Oka:2000wj, Fujiwara:2006yh}.
In this context, 
it is intuitive to expect that the 3N system is subject to extra Pauli repulsion effect,
which could be an origin of the observed short-range repulsive 3NF.
Further investigation along this line is certainly an interesting subject in future.

\begin{figure}[t]
\begin{center}
\includegraphics[width=0.78\textwidth]{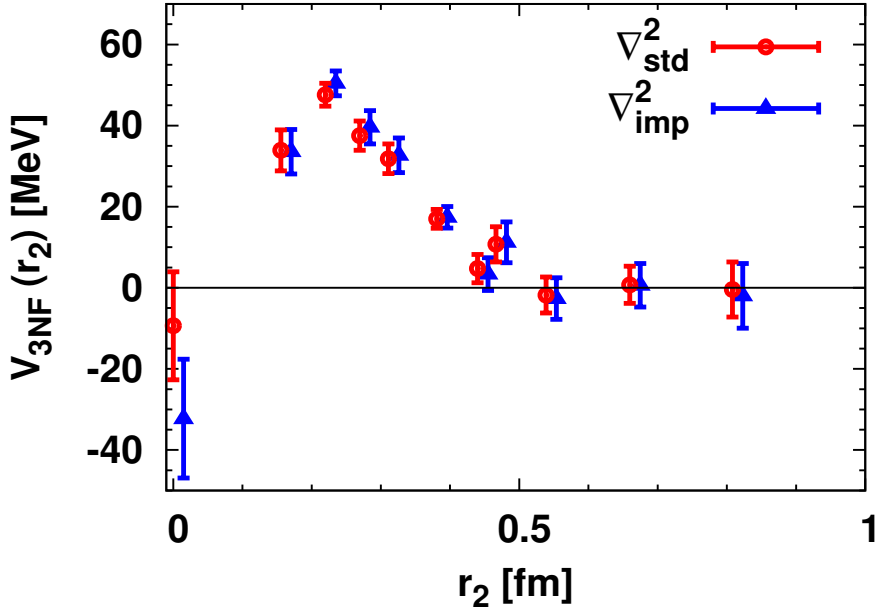}
\caption{
\label{fig:TNF_Lap}
3NF as explained in Fig.~\ref{fig:TNR}.
Circle (red) points are obtained from $\nabla^2_{\rm std}$,
and triangle (blue) points 
from $\nabla^2_{\rm imp}$.
See the text for details.
}
\end{center}
\end{figure}

Discussions on several systematic errors are in order.
First,
one may worry about the discretization error,
since the nontrivial results are obtained in the short-range part of 3NF.
In particular, 
the kinetic terms (Laplacian part in Eqs.~(\ref{eq:Sch_2N}) and (\ref{eq:Sch_3N})) 
could suffer from
a substantial effect, since they are calculated by the finite difference as
\begin{eqnarray}
\nabla^2 f(x) = 
\nabla^2_{\rm std} f(x) \equiv
\frac{1}{a^2}
\sum_{i}
\left[ f(x+a_i) + f(x-a_i) - 2 f(x) \right] .
\end{eqnarray}
In order to estimate this artifact,
we carry out additional analyses employing the improved Laplacian operator
for both of 2N and 3N systems,
\begin{eqnarray}
\nabla^2_{\rm imp} f(x)
&\equiv&
\frac{1}{12a^2}
\sum_{i}
\left[
-    ( f(x+2a_i) + f(x-2a_i) )
\right. \nn
&& \qquad\qquad
\left.
+ 16 ( f(x+a_i)  + f(x-a_i)  )
-30   f(x))
\right] . 
\end{eqnarray}
Shown in Fig.~\ref{fig:TNF_Lap}
is the comparison 
between the results of 3NF from $\nabla^2_{\rm std}$ and $\nabla^2_{\rm imp}$
at $(t-t_0)/a=8$.
We observe that 
they are consistent with each other,
and conclude that the discretization artifact of 3NF in Laplacian operator is small.
We, however, remark that this study probes only a part of discretization errors,
and an explicit calculation with a finer lattice is desirable.
Actually, the analysis with operator product expansion~\cite{Aoki:2010kx, Aoki:2010uz} shows that
2NF of $V_{2N}(r)$ tend to diverge as $r\rightarrow 0$, so 
significant discretization artifact is expected around $r=0$.
A simulation with a finer lattice is currently underway, and
we plan to report it in future publications.

Second,
the finite volume artifact could be substantial,
considering that three nucleons are accommodated in (2.5 fm)$^3$ spacial lattice box.
Such artifact, however, would mainly appear in large $r_2$ region,
and we avoid it as much as possible
by focusing on the short-range part of 3NF.
Furthermore, recall that points with relatively large $r_2$ ($r_2 \simgeq 0.5$ fm)
are carefully chosen so that they are located in off-axis directions.
We also note that the size of (2.5 fm)$^3$ is quite large 
for the heavy pion in this calculation, namely,
$m_\pi L$ is as large as $14$.
Of course, a quantitative estimate requires calculations with larger volumes,
which we defer to future studies.

Third, we study the 3NF at leading order in the derivative expansion.
Although it has been shown that higher order terms are small in 
(parity-even) 2NF~\cite{Murano:2011nz},
tiny modifications in 2NF could 
make a nonnegligible effect on 3NF.
On this point, we recall that Fig.~\ref{fig:wf} indicates
that 
the wave function
is completely dominated by 
the component with which all three nucleons are in S-wave.
Therefore, contaminations from spin-orbit forces,
which is the next leading order in the derivative expansion,
are expected to be small in this lattice setup.
It would be interesting to see how much
higher derivative terms affect 3NF
in physically lighter quark mass,
where, e.g., D-wave components get larger by larger tensor forces.

Since the lattice simulations are carried out 
only at single large quark mass point,
quark mass dependence of 3NF 
is certainly 
an important issue to be investigated. 
In the case of
2N or generalized baryon-baryon potentials,
short-range cores have the enhanced strength
and broaden range
by decreasing the quark mass%
~\cite{Aoki:2009ji, Nemura:2008sp, Inoue:2010hs, Inoue:2010es}.
We, therefore, would expect a significant quark mass dependence
exist in short-range 3NF as well.
In addition,
long-range 2$\pi$E-3NF will emerge at physically light quark masses.
As is well known, such calculation poses a significant challenge
to lattice QCD, since S/N is quickly ruined by pursuing lighter quark mass direction.
When one considers systems consisting of more nucleons, more severely this issue emerges~\cite{Lepage:1989hd}.
Theoretical developments are highly required, and some of our achievements are being
reported~\cite{Ishii:next, Inoue:2010es}.

\section{Summary}
\label{sec:summary}

We have studied three-nucleon forces (3NF) in the triton channel, $I=1/2$, $J^P= 1/2^+$,
in lattice QCD, using $N_f=2$ dynamical clover fermion 
configurations 
with the size of $16^3 \times 32$, the lattice spacing of $a=0.156$ fm
and a large quark mass corresponding to $m_\pi= 1.13$ GeV.
We have utilized the Nambu-Bethe-Salpeter wave function
to determine two-nucleon forces (2NF) and 
3NF on the same footing.
We have developed a framework to identify 3NF unambiguously,
without referring to the information of parity-odd 2NF.
When extracting 3NF,
we have fixed
the three-dimensional (3D) coordinate configuration of three nucleons.
We have proposed to employ the ``linear setup'' for the 3D-configuration,
where three nucleons are aligned linearly with equal spacings.
With this setup, 
we have shown that
the Schr\"odinger equation can be simplified to 
the $3\times 3$ coupled channel equations.
Performing lattice simulations with the linear setup,
we have found
repulsive 3NF at short distance.
Several sources of systematic errors have been discussed,
and prospects for future lattice simulations have been given as well.
In particular, 
it is important to pursue the study with lighter quark masses,
considering broad impacts of 3NF on various phenomena in nuclear physics and astrophysics,
e.g., the natures of high density matter and neutron stars.

\section*{Acknowledgements}

We thank Drs. H.~Kamada, K.~Hashimoto, S.~Ishikawa and M.~Oka
for fruitful discussions.
We are grateful for authors and maintainers of CPS++\cite{CPS},
a modified version of which is used in this study. 
We also thank  CP-PACS Collaboration
and ILDG/JLDG~\cite{conf:ildg/jldg} for providing gauge configurations.
The numerical simulations have been performed
on Blue Gene/L at KEK,
T2K at University of Tsukuba and SR16000 at YITP in Kyoto University.
This research is supported in part by MEXT Grant-in-Aid (20340047, 22540268),
Scientific Research on Innovative Areas (20105001, 20105003, 21105515),
Specially Promoted Research (13002001) and JSPS 21$\cdot$5985,
the Large Scale Simulation Program of KEK (09-23, 09/10-24) and the collaborative
interdisciplinary program at T2K-Tsukuba (09a-11, 10a-19).

\appendix

\section{Effective two-nucleon forces in three-nucleon systems}
\label{sec:eff_2N}

We investigate the Schr\"odinger equation for the 3N system, Eq.~(\ref{eq:Sch_3N}),
from a 
viewpoint of
effective 2NF.
More specifically,
we take the summation over the 
location of a spectator nucleon $N(\vec{x}_3)$,
\begin{eqnarray}
\psi_{\rm eff}(\vec{r}) \equiv 
\sum_{\vec{x}_3} \psi_{3N}(\vec{r},\vec{\rho}) = 
\sum_{\vec{\rho}} \psi_{3N}(\vec{r},\vec{\rho}),
\label{eq:wf_eff_2N}
\end{eqnarray}
and define the effective potential 
between $N(\vec{x}_1)$ and $N(\vec{x}_2)$
through 
the 
following 
effective Schr\"odinger equation,
\begin{eqnarray}
\biggl[ 
- \frac{1}{2\mu_r} \nabla^2_{r}
+ V_{\rm eff} (\vec{r})
\biggr] \psi_{\rm eff}(\vec{r})
= E_{3N} \psi_{\rm eff}(\vec{r}) .
\label{eq:Sch_eff_2N}
\end{eqnarray}
%

Since
the DoF of $\vec{\rho}$ is integrated out beforehand,
computational cost is reduced drastically.
This summation is also expected to enhance the S/N.
Due to these features,
effective 2NF serve as a suitable framework
for preceding simulations to much more expensive 
linear setup simulations (or any other fixed 3D-configuration simulations.)
It can also determine $E_{3N}$, which provides
an $r_2$-independent shift of 3NF in the linear setup simulations.

As the 3N system, we consider the triton channel,
$I=1/2$, $J^P= 1/2^+$.
Because the spectator nucleon is projected to S-wave,
possible quantum numbers between the (effective) 2N
are only $^{2S+1}L_J =$ $^1S_0$, $^3S_1$, $^3D_1$.
We calculate the effective 2NF $V_{\rm eff}^{IS}$,
i.e.,
the central $V_{C,\rm eff}^{10}$,
$V_{C,\rm eff}^{01}$
and the tensor $V_{T,\rm eff}^{01}$
potentials,
where $I$ and $S$ denote
the isospin and spin, respectively.

\begin{figure}[tb]
\begin{minipage}{0.48\textwidth}
\begin{center}
\includegraphics[width=1.0\textwidth]{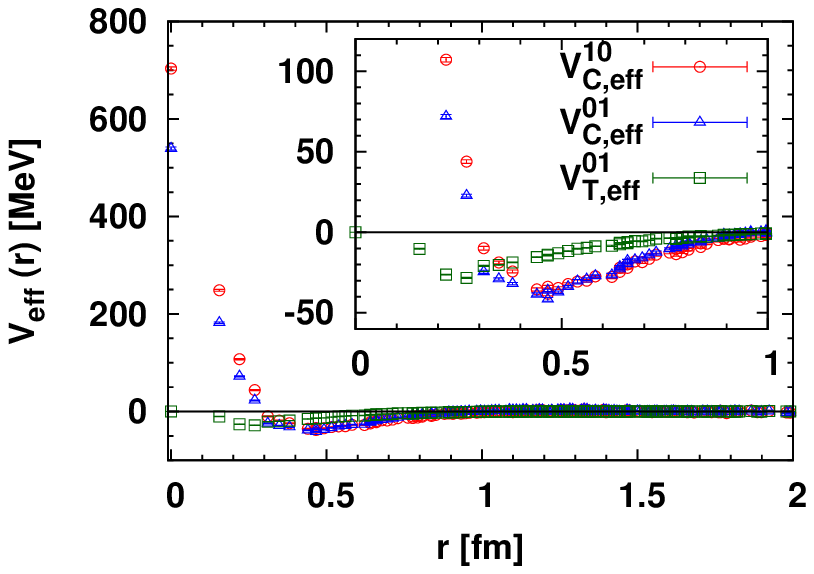}
\caption[hoge]{
\label{fig:pot_3b.eff_2N}
Same as Fig.~\ref{fig:pot_2N},
except that the
effective 2NF in the triton channel
are plotted,
instead of the genuine 2NF.
}
\end{center}
\end{minipage}
\hfill
\begin{minipage}{0.48\textwidth}
\begin{center}
\includegraphics[width=1.0\textwidth]{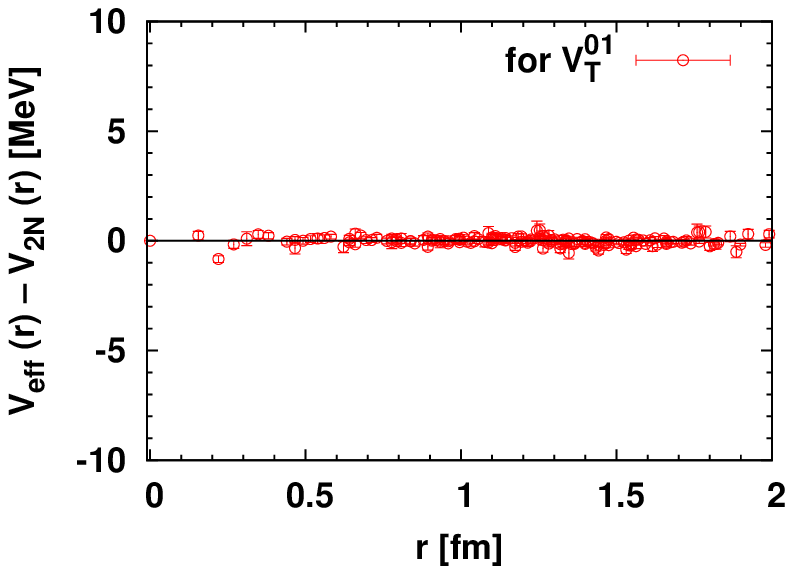}
\caption[hoge]{
\label{fig:pot_diff.ten}
The difference between the effective 2NF 
and the genuine 2NF
for the tensor potential $V_{T}^{01}$.
}
\end{center}
\end{minipage}
\end{figure}

Lattice simulations are carried out with the same setup 
as in Section~\ref{sec:setup:linear},
except for that the measurement is taken at
16 source time slices for each configuration,
and with wider sink time range of $2 \leq (t-t_0)/a \leq 11$.
In Fig.~\ref{fig:pot_3b.eff_2N}, 
we show results for 
$V_{\rm eff}^{IS}(r)$
obtained at $(t-t_0)/a=8$.
The results agree with those from different $t$ as long as $(t-t_0)/a \geq 7$.
The $r$-independent shift by the energy in Eq.~(\ref{eq:Sch_eff_2N})
is determined from the long range behavior of the potential
(the method called ``from $V$'' in Ref.~\citen{Aoki:2005uf}).
We obtain $E_{3N} \simeq -3$~MeV
from both of $V_{C,\rm eff}^{10}$ and $V_{C,\rm eff}^{01}$, consistently.
Note that the tensor potential does not suffer from
this overall shift~\cite{Aoki:2009ji}.

It would be of interest to compare
effective 2NF $V_{\rm eff}(r)$ and 
genuine 2NF $V_{2N}(r)$
and extract the effect of the 3N system.
Some of the effects are attributed 
to genuine 2NF with nontrivial 3N correlations,
and the others are originated by genuine 3NF.
In this way, we can indirectly access the effect of 3NF.

In Fig.~\ref{fig:pot_diff.ten},
we plot $V_{\rm eff}(r) - V_{2N}(r)$ 
for the tensor potential, $V_T^{01}$.
One can see that the difference is
consistent with zero,
and there is no indication of the 3NF effect.
The results of $V_{\rm eff}(r) - V_{2N}(r)$ 
for the central potentials $V_C^{10}$ and $V_C^{01}$ 
are similarly consistent with zero,
with larger statistical errors than the case of $V_T^{01}$
by a factor of 2-4.

This (non-)observation can be explained as follows.
First of all, 
the 2$\pi$E-3NF component is expected to be suppressed
with $m_\pi = 1.13$ GeV in this calculation.
Furthermore, $\Delta$-excitation is suppressed by
the S-wave projection on a spectator nucleon.
We still have short-range repulsive 3NF observed 
in Section~\ref{sec:results:linear}.
It turns out, however, that 
such 3NF effect is obscured due to the volume factor
in the summation of Eq.~(\ref{eq:wf_eff_2N}).
For instance, if we consider effective 2NF
with the distance of $r \simeq 0.2$ fm, 
a spectator nucleon is involved in 3NF
only when it resides within $\simleq 0.4$ fm 
from the center of the 2N of concern.
Therefore, the relative volume factor
amounts to 
$\simleq \frac{4}{3}\pi (0.4\,{\rm fm})^3 / (2.5\,{\rm fm})^3 = 0.017$ and
thus the 3NF affect effective 2NF only  $\ll 1$ MeV.
This magnitude is comparable to the size of the statistical errors in the effective 2N study.

These discussions in turn show that
it is essential to perform the simulation with fixed 3D-configuration of 3N
in order to extract 3NF.

\end{document}